\documentclass[conference]{IEEEtran}

\usepackage{booktabs}
\usepackage{xcolor}
\usepackage{graphicx}
\usepackage{float}
\usepackage{cite}
\usepackage{url}
\usepackage{hyperref}

\usepackage{listings}
\usepackage{enumitem}
\newcommand{\OurOrg}{Microsoft}
\newcommand{\OurStorage}{OneDrive}
\newcommand{\AnonAuthors}{Nancy Baym, Jenna Butler, Sankeerti Haniyur,
    Emerson Murphy-Hill, Constance Noonan Hadley, Jina Suh}

\newcommand{\NParticipants}{447}
\newcommand{\NReviews}{1,788}

\newcommand{\PctLevelOne}{10.1}
\newcommand{\PctLevelTwo}{34.9}
\newcommand{\PctSenior}{31.1}
\newcommand{\PctPrincipalPlus}{23.9}
\newcommand{\PctMale}{75.6}
\newcommand{\PctFemale}{18.6}
\newcommand{\PctGenderOther}{5.8}
\newcommand{\PctReviewMultiplePerDay}{27.3}
\newcommand{\PctReviewDaily}{23.9}
\newcommand{\PctReviewFewTimesWeek}{36.5}
\newcommand{\PctReviewWeekly}{6.5}
\newcommand{\PctReviewEveryFewWeeks}{5.8}
\newcommand{\MeanReviewConfidence}{3.61}
\newcommand{\PctAIGenNever}{2.2}
\newcommand{\PctAIGenOccasionally}{18.1}
\newcommand{\PctAIGenSometimes}{24.4}
\newcommand{\PctAIGenOften}{42.1}
\newcommand{\PctAIGenAlways}{13.2}
\newcommand{\PctAIGenAnyUse}{97.8}
\newcommand{\PctAIReviewNever}{18.8}
\newcommand{\PctAIReviewOccasionally}{28.0}
\newcommand{\PctAIReviewSometimes}{19.0}
\newcommand{\PctAIReviewOften}{24.6}
\newcommand{\PctAIReviewAlways}{9.6}
\newcommand{\PctAIReviewAnyUse}{81.2}
\newcommand{\MeanComfortGen}{6.17}
\newcommand{\MeanComfortReview}{5.97}

\newcommand{\NCommentRequired}{804}
\newcommand{\PctCommentRequired}{45.0}

\newcommand{\AICompCoef}{-0.034}
\newcommand{\AICompCIlo}{-0.130}
\newcommand{\AICompCIhi}{+0.061}
\newcommand{\AICompP}{p=0.481}
\newcommand{\AICompD}{-0.07}

\newcommand{\PowerWithin}{>99}
\newcommand{\GaiD}{-0.31}

\newcommand{\GaiDInOurUnits}{-0.28}

\newcommand{\ComfortDMin}{0.28}
\newcommand{\ComfortDMax}{0.30}

\newcommand{\SrCompCoef}{+0.178}

\newcommand{\SrCompP}{p<0.001}
\newcommand{\SrEffCoef}{+0.148}

\newcommand{\SrEffP}{p=0.001}

\newcommand{\JunLazyCoef}{+0.084}

\newcommand{\JunLazyP}{p=0.075}
\newcommand{\JunQualCoef}{-0.184}

\newcommand{\JunQualP}{p=0.002}
\newcommand{\JunReadCoef}{-0.142}

\newcommand{\JunReadP}{p=0.019}
\newcommand{\JunSecCoef}{-0.123}

\newcommand{\JunSecP}{p=0.026}

\newcommand{\JunSkillCoef}{-0.241}

\newcommand{\JunSkillP}{p=0.003}
\newcommand{\JunMastCoef}{-0.388}

\newcommand{\JunMastP}{p<0.001}

\newcommand{\JunGiveUpCoef}{+0.170}

\newcommand{\JunGiveUpP}{p=0.021}

\newcommand{\MCPassRate}{48.1}
\newcommand{\MCRhoObs}{+0.414}
\newcommand{\MCRhoExp}{+0.085}
\newcommand{\MCRhoDiff}{+0.330}

\newcommand{\ExclSrCompP}{p=0.006}
\newcommand{\ExclSrEffP}{p=0.027}
\newcommand{\CommentOptionalP}{p=0.559}
\newcommand{\CommentRequiredP}{p=0.260}
\newcommand{\AIMentionLabeled}{28}
\newcommand{\AIMentionLabeledNonempty}{238}
\newcommand{\AIMentionUnlabeled}{7}
\newcommand{\AIMentionUnlabeledNonempty}{236}
\newcommand{\AIMentionOR}{4.36}
\newcommand{\AIMentionP}{p<0.001}
\newcommand{\RobustSrCompP}{p=0.004}
\newcommand{\RobustSrEffP}{p=0.007}

\newcommand{\NExclReviews}{1,341}
\newcommand{\ExclAICompCoef}{-0.018}
\newcommand{\ExclAICompP}{p=0.759}
\newcommand{\ExclAIEffCoef}{+0.019}
\newcommand{\ExclAIEffP}{p=0.734}
\newcommand{\ExclAILazyCoef}{+0.064}
\newcommand{\ExclAILazyP}{p=0.243}

\begin{document}

\title{Where AI Use Is Normalized, Author Seniority but Not AI Disclosure Biases Code Review}

\author{%
\IEEEauthorblockN{%
  Jenna Butler\IEEEauthorrefmark{1},
  Emerson Murphy-Hill\IEEEauthorrefmark{1},
  Constance Hadley\IEEEauthorrefmark{2},
  Alexa Tipton\IEEEauthorrefmark{1},\\
  Nancy Baym\IEEEauthorrefmark{1},
    Sankeerti Haniyur\IEEEauthorrefmark{4},
    Jina Suh\IEEEauthorrefmark{3}}
\IEEEauthorblockA{%
  \IEEEauthorrefmark{1}Microsoft, USA\quad
  \IEEEauthorrefmark{2}Institute for Life at Work, USA\quad
    \IEEEauthorrefmark{3}University of Washington, USA\quad
    \IEEEauthorrefmark{4}Independent Researcher\\
    \{jennbu, emerson.rex, alexatipton, baym\}@microsoft.com,\\
    cnhadley@institutelifework.org,
    skhaniyur@gmail.com,
    jinasuh@cs.washington.edu}}

\maketitle

\begin{abstract}
Code-review tools increasingly display whether an author used AI when preparing
pull requests. 
Recent
studies report that AI disclosure 
makes reviewers rate otherwise
identical work as less competent, with the penalty falling hardest on
already-marginalized authors. In a within-subjects experiment in an
AI-normalized organization, \NParticipants\ software engineers each
reviewed the same four code snippets under an AI-use disclosure embedded
in the commit message and author-seniority labels. AI disclosure did not
bias perceptions of code effectiveness or author competence, whereas a
seniority label significantly biased both. Together, these findings show
no detected penalty for disclosed AI use in this setting, while
confirming that the author information shown in code review can still
bias evaluations.
\end{abstract}

\begin{IEEEkeywords}
code review, AI disclosure, status bias
\end{IEEEkeywords}

\section{Introduction}

\begin{figure}[!t]
  \centering
  \includegraphics[width=\columnwidth]{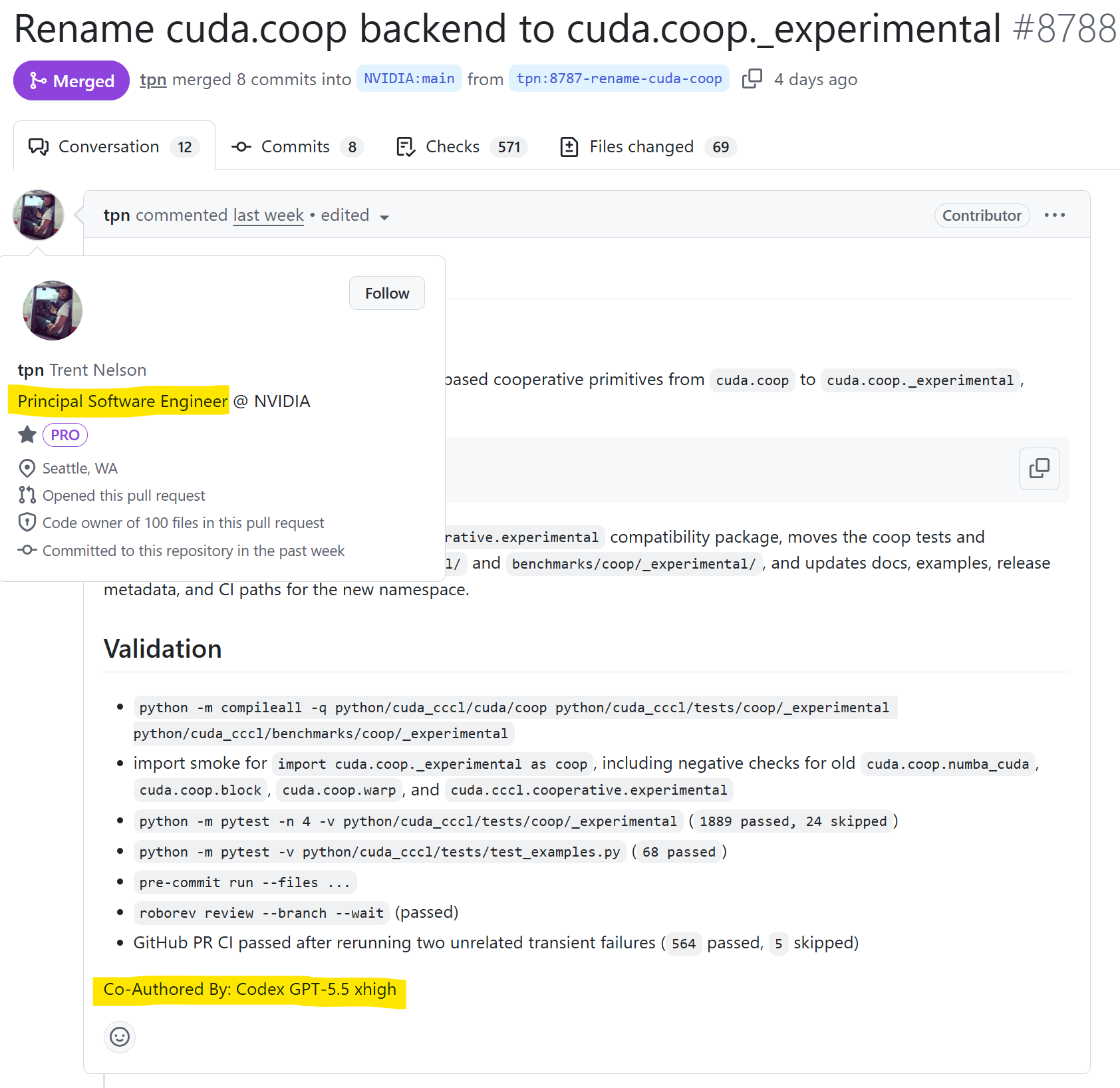}
    \caption{An open source pull request~\cite{nvidia8788} identifying the author's seniority and that AI was used. Highlighting added to aid the reader.}
  \label{fig:senior-ai-example}
\end{figure}

When a developer opens a code review, the tool shows them the code
alongside labels about its author -- name, photo, title, and,
increasingly, whether the author used AI
assistance.
An example of such labels is shown in Figure~\ref{fig:senior-ai-example}.
Do those labels bias how reviewers judge the code or the coder?

Three recent studies say yes, and single out the AI label.
Schilke and Reimann~\cite{schilke2025} find that people who disclose AI use are trusted less than
people who do not, projecting this penalty will persist even as AI diffuses.
Inside a large technology company, Gai and
colleagues~\cite{gai2025} report a similar overall AI usage penalty in code review,
but note it was driven entirely by non-adopters.
An earlier study by Huang and colleagues~\cite{huang2020} found that
software engineers rejected machine-labeled patches at higher rates
than human-labeled ones.
Taken together, these results warn that the AI label is a new source
of bias in software engineering, one that may compound inequalities
for already-marginalized developers.

Those warnings come from populations in
which AI use itself is arguably the unfamiliar variable. We ask whether the
penalty survives in a population where AI use is normalized, and
whether a different author label code-review tools have long surfaced --
author seniority -- carries bias of its own. \NParticipants\
software engineers each reviewed the same four code snippets in a
within-subjects 2$\times$2 design that randomized, per snippet, the
author's seniority (junior vs.\ senior) and AI-use disclosure (used AI
vs.\ did not). The code itself was held constant across these conditions. Across
perceptions of code effectiveness and author competence,
the AI label did not bias either while the seniority label significantly
biased both, with no seniority\(\times\)AI interaction. 
We do not dispute the effects identified in prior work, 
but instead argue they are context-dependent. 
In a future where AI usage is normalized, penalties for disclosure may fade,
 whereas biases associated with attributes such as gender and seniority are, 
 unfortunately, likely to persist.

\section{Related Work}

\medskip\noindent\textbf{The three AI-penalty results we test against.}
\label{sec:related-ai}
The Introduction names three prior studies our design is powered to
replicate; we summarize the evidence for the comparison in
Section~\ref{sec:why-no-penalty}. Schilke and
Reimann~\cite{schilke2025} report thirteen pre-registered experiments,
from analytics to creative writing, in which the AI-disclosure trust
penalty persisted even when AI was framed only as proofreading. Their
meta-analysis finds the penalty attenuates but does not disappear
among technology-favorable evaluators, suggesting persistence as AI
diffuses. Gai and colleagues~\cite{gai2025}, studying a large Chinese
technology company, found AI-assisted engineers rated 9\% less
competent ($d=\GaiD$), with the penalty driven by non-adopter
reviewers ($d=-0.46$ vs.\ $d=-0.02$ among adopters). Huang and
colleagues~\cite{huang2020} ran a pre-LLM fMRI study in which software
engineers acknowledged an anti-machine code-review bias roughly three
times their gender bias and rejected machine-labeled patches more
often than human-labeled ones; there, the machine was fully
autonomous, not a human assistant.

\medskip\noindent\textbf{Demographic cues in code review.}
Long before AI disclosure, code review
was already known to be sensitive to who the author appeared to be.
Terrell and colleagues~\cite{terrell2017} found that women's pull requests on
GitHub were accepted at higher rates than men's when author gender
was not visible, and at lower rates when it was. Murphy-Hill and
colleagues~\cite{murphyhill2022} found that authors who were women,
non-White, or older received more pushback than men, White, or younger
authors across more than two million Google code reviews, controlling
for seniority and tenure. Yabesi and colleagues~\cite{yabesi2024} found that
author seniority did not influence code quality
assessments but did affect how long the reviewer spent on the PR.
Moreover, eye-tracking studies have found that programmers look at social and 
identifying signals on a pull request significantly more than they self-report 
(with very large effect sizes, such as 100\% of participants fixating on 
avatar images despite only 2\% reporting doing so),
showing that developers are often unaware of the extent to 
which they process these cues~\cite{ford2019}. The
AI-use label is the newest addition to a longer list of author
attributes a review tool \emph{may} surface; our study adds the
seniority label to the experimental side of that list.

\medskip\noindent\textbf{Status penalty effects.}
Our seniority prediction comes from older results outside
software engineering. Status characteristics theory~\cite{berger1972}
holds that people defer to globally valued attributes, such as
organizational rank, when evaluating task-specific performance, even
when the attribute carries no diagnostic information about the task.
Bunderson~\cite{bunderson2003} showed how, given ambiguous
or minimal information, groups use status cues to ascribe higher or
lower expertise to fellow members. In decentralized groups
with less direct knowledge of teammates, employees tended to use
technical certification level and tenure as proxies for expertise.
Our study environment similarly provided little direct information
about the coder's capabilities, so participants likely
defaulted to seniority to
ascribe expertise. The status penalty effect (or its opposite, the
status halo effect)~\cite{nisbett1977} generalizes the claim:
unfavorable global impressions bleed into ratings of specific
behaviors. Together these accounts predict that a more junior status
label attached to identical code should undermine ratings of that
code and the author who created it.

\section{Method}

We ask two research questions:
\begin{itemize}
  \item \textbf{RQ1 (AI label).} Does the AI-use label bias code-review
  ratings?
  \item \textbf{RQ2 (Seniority label).} Does the seniority label bias
  code-review ratings?
\end{itemize}

\subsection{Design}
\begin{figure}[t]
  \centering
  \includegraphics[width=\columnwidth]{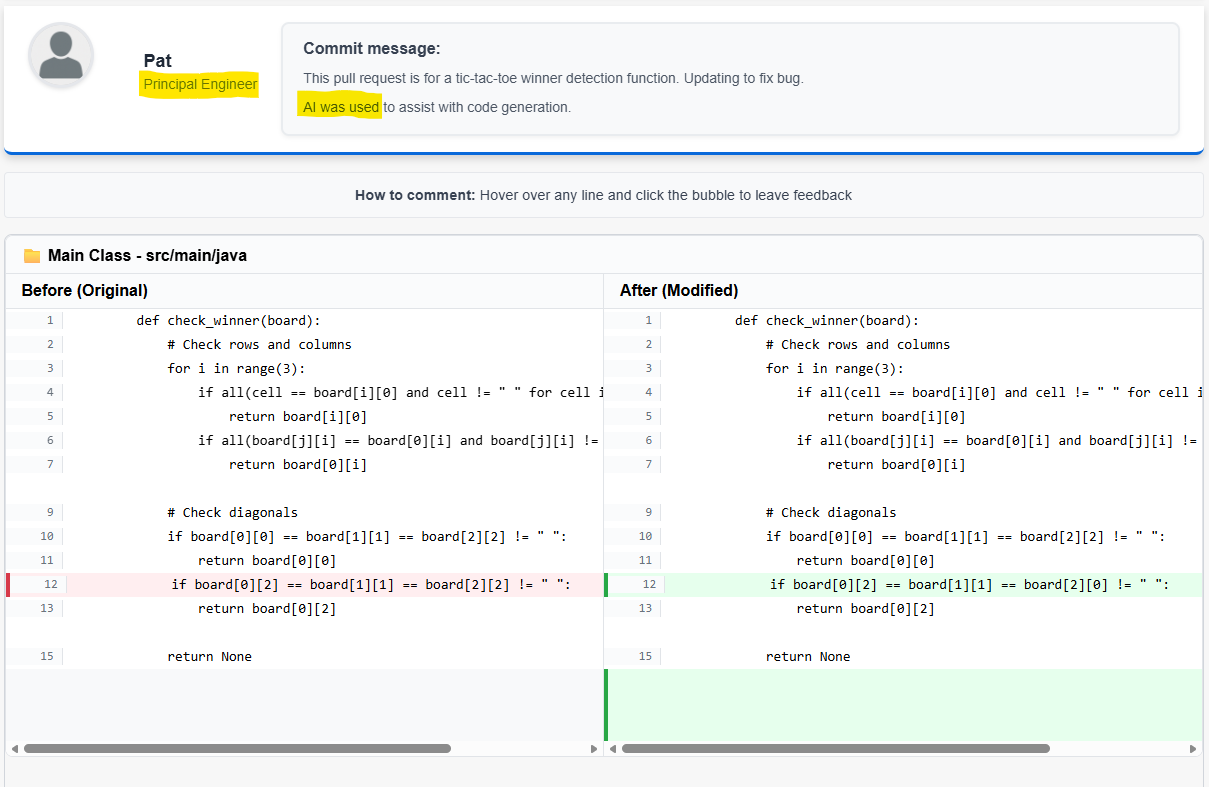}
    \caption{Example of code review website used in the study, displaying the purpose of the code, the seniority of the coder, and AI use disclosure. Highlighting added to aid the reader and was not shown to participants.
  }
  \label{fig:example}
\end{figure}

Each participant reviewed four code snippets in a within-subjects
2$\times$2 factorial: author title (Level 1
vs.\ Principal) crossed with AI disclosure (``used AI'' vs.\ ``did not
use AI'').
Note that these two levels are not adjacent, so a single-level difference 
would likely yield a smaller effect.
Assignment of conditions to code snippets was randomized per
participant. Four code snippets--held fixed across conditions--spanned two programming languages and varying numbers of seeded bugs per
snippet. Three were drawn from prior code review studies by Spadini and colleagues:
one (4 seeded bugs) is from their test-driven code review study~\cite{spadini2019tdcr},
and two (2 seeded bugs each) are from their study on the effects of existing review comments~\cite{spadini2020primers}.
The fourth snippet (TicTacToe; no seeded bugs) was AI-generated for this study to add heterogeneity.
The full snippets are reproduced in the supplementary material, Section~\ref{app:snippets}.
Figure~\ref{fig:example} shows an example of what participants saw.
The initial user interface was piloted with four developers to ensure a reasonable user experience,
resulting in refinements to make it look more similar to existing code review tools.  

\subsection{Participants}
\label{sec:participants}
\NParticipants\ full-time software engineers at \OurOrg{}
completed the study between October~13 and November~10, 2025,
producing \NReviews\ reviews. Seniority of
participants -- an attribute orthogonal to the manipulated seniority of the purported code author -- ranged as follows: Level 1 (\PctLevelOne\%),
Level 2 (\PctLevelTwo\%), Senior Engineer (\PctSenior\%), and
Principal Engineer or Above (\PctPrincipalPlus\%). The majority were male
(\PctMale\%), with \PctFemale\% female and \PctGenderOther\% preferring not to answer or
missing data. Participants were highly experienced conducting code
reviews, undertaking them every few weeks
(\PctReviewEveryFewWeeks\%), weekly (\PctReviewWeekly\%), a few times a week (\PctReviewFewTimesWeek\%), daily
(\PctReviewDaily\%), or multiple times a day (\PctReviewMultiplePerDay\%). They averaged \MeanReviewConfidence\ in
confidence in conducting code reviews on a 5-point confidence scale
from not at all confident to extremely confident.

\subsection{Dependent Variables (DVs)}
After each code review, participants answered a set of questions
about the code and its author, plus one decision question. We grouped
items by what they asked about; the verbatim wording is reproduced below
(the full instrument is in the supplementary material, Section~\ref{app:survey}). 
Our primary outcomes are three $z$-scored composites
defined a priori from these item groupings: \textsc{CodeEffectiveness},
\textsc{CoderCompetence}, and \textsc{CoderLaziness}. We also report
their constituent items, the hiring recommendation, and the number of
comments as secondary outcomes.

\medskip\noindent\textbf{Code ratings.}
Three items rated on a 5-point scale from Very poor to Excellent.
We $z$-score each and average them to form the
\textsc{CodeEffectiveness} composite.
\begin{itemize}[leftmargin=*,nosep]
  \item \emph{Quality.} ``Overall, how would you rate the quality of
        this code?''
  \item \emph{Readability.} ``How would you rate the readability of
        this code?''
  \item \emph{Security.} ``How would you rate the security of this code?''
\end{itemize}

\medskip\noindent\textbf{Author capability.}
A single item on the same 5-point scale as the code ratings (``How would you rate coding capabilities of this author?''), $z$-scored
and combined with the competence traits below to form the
\textsc{CoderCompetence} composite.

\medskip\noindent\textbf{Competence traits.}
Three adjectives rated on a 7-point Likert scale from Strongly disagree
to Strongly agree, $z$-scored and averaged with Author capability above
to complete the \textsc{CoderCompetence} composite.
Prompt: ``Please rate the extent to which you believe the following
adjectives or phrases accurately describe the author of this code'' (Competent, Skillful, Masterful).
These items were inspired by the adjectives used to measure perceived
competence in Reif and colleagues'~\cite{reif2025} study of social
evaluation penalties for AI use.

\medskip\noindent\textbf{Laziness traits.}
Three adjectives, same prompt and 7-point scale as the competence
traits, $z$-scored and averaged to form the \textsc{CoderLaziness}
composite (Irresponsible, Lazy, Gives up easily).
These items were likewise inspired by the laziness and diligence
adjectives in Reif and colleagues~\cite{reif2025}.

\medskip\noindent\textbf{Hiring recommendation.}
A single item on a 5-point scale from Very unlikely to Very likely,
analyzed as a stand-alone DV: ``Imagine you are a hiring manager responsible for recruiting a
software engineer at your company. Based on the code you just
reviewed, how likely is it that you would recommend the author
to be considered for the position?''

\medskip\noindent\textbf{Number of comments.}
A behavioral measure, analyzed as a stand-alone DV: the count of
inline comments the reviewer left on the diff, log-transformed.

\medskip Leaving an inline comment was initially optional, but because
some participants left no comments even on snippets with seeded bugs,
we made at least one comment required for participants who submitted
later in data collection (\NCommentRequired\ reviews,
\PctCommentRequired\%); we include this regime change as a covariate.

\subsection{Analysis}
For each dependent variable (DV) we fit a linear mixed-effects model with participant as a
random intercept:
\[
\mathrm{DV} \sim \mathrm{isSenior} \times \mathrm{usedAI}
  + C(\mathrm{codeID}) + \mathrm{commentRequired}
\]
where \textsf{isSenior} is 1 for author Principal titles, \textsf{usedAI} is 1
when AI use was disclosed, \textsf{codeID} controls for snippet, and
\textsf{commentRequired} controls for the comment-requirement regime
change mid-collection. We report coefficients, 95\% confidence intervals,
and $p$-values. Because the study was not pre-registered, we treat the
three composites as the primary outcomes and the individual-item and
robustness-model analyses as secondary. A secondary model adds reviewer seniority
(\textsf{reviewerIsSenior}) and its interactions as a robustness check
against an in-group-favoritism interpretation.

\section{Results}

\subsection{RQ1: No AI Penalty
}
\begin{figure*}[t]
  \centering
  \includegraphics[width=\textwidth]{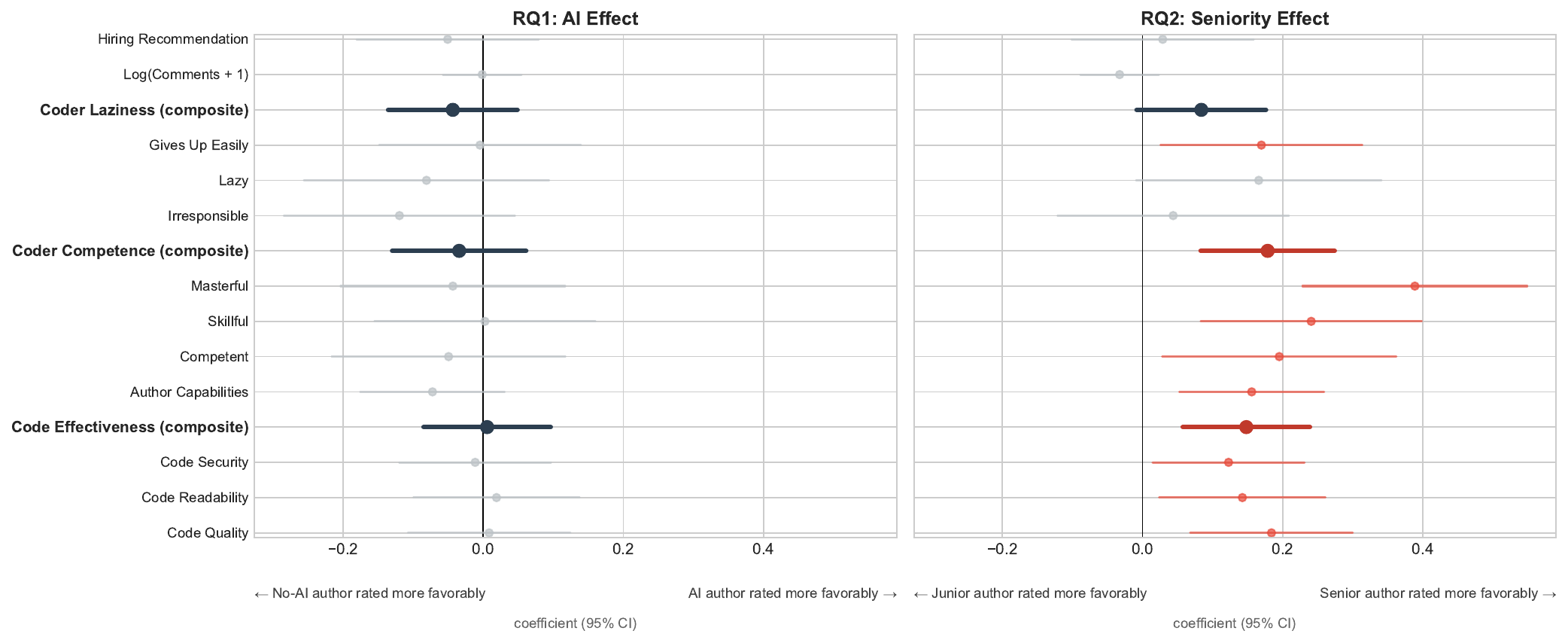}
  \caption{Mixed-effects coefficients (95\% CI) for the AI effect (left)
  and the seniority effect (right) on a shared coefficient scale,
  across all DVs. Bold rows are composites; light rows are their
  constituent items. Red bars indicate $p<.05$. For laziness items
  (\textsc{CoderLaziness} and its components), coefficient signs are
  flipped so that, in both panels, points to the right of zero always
  mean the cued group (AI author / Senior author) was rated more
  favorably.}
  \label{fig:forest}
\end{figure*}

Labeling code as having involved AI had no detectable effect on any
of the measured outcomes. AI use did not bias ratings on
\textsc{CoderCompetence}; the AI coefficient was
\(\AICompCoef\) (95\% CI \([\AICompCIlo, \AICompCIhi]\), \(\AICompP\), \(d=\AICompD\)).
All dependent variables in the forest plot
(Figure~\ref{fig:forest}) are consistent with a null AI effect.
Furthermore, there was no measurable Seniority\(\times\)AI
interaction effect.

Several robustness checks support our conclusion that participants did
not penalize authors for using AI
(Appendix~\ref{app:robustness}). First, spontaneous AI mentions were
more common in feedback on AI-labeled than non-AI-labeled reviews
(\AIMentionLabeled/\AIMentionLabeledNonempty\ vs.\
\AIMentionUnlabeled/\AIMentionUnlabeledNonempty; \(\AIMentionP\)),
providing evidence that at least some participants noticed the
disclosure. Second, the study had \(\PowerWithin\%\) power to detect
the AI penalty reported by Gai and colleagues, and our confidence
interval excludes an effect of that magnitude. Finally, excluding the
AI-generated TicTacToe snippet did not change the result.

This lack of penalty likely reflects the fact that our participants are highly
accustomed to AI tools:
\begin{itemize}[leftmargin=*]
  \item \textbf{High personal usage.} Almost all (\PctAIGenAnyUse\%) reported using AI themselves to generate
        code (response distribution: never (\PctAIGenNever\%); occasionally (\PctAIGenOccasionally\%);
        sometimes (\PctAIGenSometimes\%); often (\PctAIGenOften\%); always (\PctAIGenAlways\%)). Most (\PctAIReviewAnyUse\%)
        use AI for code reviews as well (never: \PctAIReviewNever\%; occasionally: \PctAIReviewOccasionally\%;
        sometimes: \PctAIReviewSometimes\%; often: \PctAIReviewOften\%; always: \PctAIReviewAlways\%).
  \item \textbf{High disclosure comfort.} Participants reported near-ceiling comfort disclosing their AI use for both code generation
        (\MeanComfortGen\ on a 7-point scale) and code review (\MeanComfortReview).
        Their comfort with others knowing about their AI use modestly exceeded their active desire for others to know 
        (\(d\approx \ComfortDMin\)--\(\ComfortDMax\)), but scores for both remained very high across all manager and colleague contexts.
\end{itemize}
Consistent with the adopter moderation reported
by Gai and colleagues---who found the penalty vanished among AI-adopting reviewers---an AI penalty 
does not appear in this highly AI-normalized population.

\subsection{RQ2: A Junior Penalty}
\label{sec:rq2}
In contrast to the lack of effect of AI, seniority did bias
judgements of both the code and the coder. Junior authors were
penalized even though the evaluated code was held constant. On
\textsc{CoderCompetence}, the senior-vs-junior coefficient was
\(\SrCompCoef\) (\(\SrCompP\)). On \textsc{CodeEffectiveness} it was
\(\SrEffCoef\) (\(\SrEffP\)). Code from junior authors
was rated as lower quality (\(\JunQualCoef\), \(\JunQualP\)), less readable
(\(\JunReadCoef\), \(\JunReadP\)), and less secure (\(\JunSecCoef\),
\(\JunSecP\)). Junior authors were seen as less masterful
(\(\JunMastCoef\), \(\JunMastP\)), less skillful (\(\JunSkillCoef\),
\(\JunSkillP\)) and as more likely to give up easily
(\(\JunGiveUpCoef\), \(\JunGiveUpP\)). The broader
\textsc{CoderLaziness} composite trended in the same direction but was
not significant (\(\JunLazyCoef\), \(\JunLazyP\)).
Neither hiring recommendations nor reviewer comment counts were
affected by seniority.

The junior penalty was also robust to alternative specifications
(Appendix~\ref{app:robustness}). The seniority effects on
\textsc{CoderCompetence} and \textsc{CodeEffectiveness} remained
significant after excluding TicTacToe and after adding reviewer
seniority and its interactions. The author-seniority effect did not
differ significantly by reviewer seniority, suggesting that it was
not driven by senior reviewers favoring senior authors. The null
effect on comment count also held when the optional- and
required-comment regimes were analyzed separately.

\section{Discussion}

\subsection{Why the AI-Disclosure Penalty Does Not Appear}
\label{sec:why-no-penalty}
Given that our confidence interval excludes the magnitude of Gai and
colleagues' competence penalty, why did the penalty not appear here?
We have two plausible explanations.

\emph{AI-naive vs.\ AI-normalized evaluators.}
Schilke and Reimann read their attenuator analysis as evidence the AI penalty will
persist as adoption increases. However, Gai and colleagues found that code
reviewers who had actively adopted AI showed no penalty ($d=-0.02$), though their
overall population did because adoption was only 41\%. By surveying a professional
population where AI use is normalized (\PctAIGenAnyUse\% used AI for code generation
and more than 80\% used it for code reviews), we essentially put Schilke and Reimann's
pessimistic projection against Gai's adopter sub-group finding. Our results align with
the latter in the scoped setting we tested. Participants also expressed confidence disclosing their AI usage
to managers and colleagues, indicating a low stigma attached to doing so.
Thus, among a population in which AI use is normalized culturally and
operationally, we do not detect an AI penalty under the commit-message
disclosure tested here.

\emph{Autonomous agent vs.\ tool use.} Huang and colleagues framed the
AI condition as fully autonomous: a bot
produces the patch without human intervention and the reviewer evaluates it. 
Our disclosure names
the human as author and the AI as an assistive tool, leaving the human
accountable for the submitted code. Schilke and Reimann further argue
that humans who disclose AI use are trusted \emph{less} than
autonomous AI agents performing the same task because co-authorship
blurs the locus of agency~(their H4). Our data do not reproduce that
pattern: framing the author as a tool user did not trigger a measurable
legitimacy discount in this population. Locus-of-agency ambiguity may
matter more for evaluators who are themselves uncertain about AI's role
in professional work; evaluators who use AI tools daily may not
experience tool-use disclosure as a deviation from normative
expectations.

\subsection{Bias Has Not Disappeared: The Junior Penalty}

Our finding of a pervasive junior penalty stands in direct contrast to Yabesi and colleagues~\cite{yabesi2024}, 
who found no effect of author seniority on explicit code quality assessments. 
A likely explanation for this divergence is the participant population: 
Yabesi and colleagues studied university students with minimal professional experience, 
whereas our sample comprised professionals. 
Professional engineers are steeped in an organizational hierarchy where positional status 
is a highly salient cue, making them more susceptible to using job seniority as 
a proxy for expertise.

As shown in Bunderson~\cite{bunderson2003}, individuals may default
to indirect cues, such as author status, when evaluating the
expertise of others---especially in low-context situations. Our
experiment asked participants to evaluate the competence and
laziness of code authors in a scenario that provided little
diagnostic information beyond the code. Both status
characteristics theory~\cite{berger1972} and classic halo
work~\cite{nisbett1977} would predict that, in the absence of other
relevant information, job seniority would be used as a proxy for
expertise and effort. Consistent with these theories, we find that
junior status cues were associated with lower ratings of code
quality, readability, and security, even though the underlying code
was the same. Moreover, the purported code authors paid
the price for their junior status designation, receiving lower
ratings of competence and being rated as more likely to give up easily
than the senior status code authors, although the broader laziness
composite was not significantly different. The junior penalty was consistent regardless of
the seniority of the participant, indicating this was a universal
bias in our sample.


\subsection{How Much of the Field-Observed Junior Penalty Is Bias?}
\label{sec:decompose}
Murphy-Hill and colleagues~\cite{murphyhill2022} analyzed more than two
million Google code reviews and found that, controlling for tenure
and properties of the change, senior engineers faced roughly $62\%$
lower odds of high pushback than junior engineers. Broadly, that
difference could come from either the author or the reviewer: senior
engineers' code and tasks may genuinely differ from junior
engineers', or reviewers may evaluate identical work differently
when they know the author's seniority. The first is only partially
controlled away in observational data; the second cannot be tested
at all without holding code and task constant.

Our experiment can tease apart these two explanations: it holds code and
task constant and varies only the displayed
title, so any movement in our ratings is the second explanation by
construction. As an illustrative order-of-magnitude comparison, putting
the two studies on a common scale (details in the supplementary
material, Section~\ref{app:decompose}) suggests that bias on identical
code could account for roughly $33\%$ of this field difference. This
estimate rests on cross-study scaling between a vignette composite and a
field-observed pushback outcome, so it should not be read as an exact
point estimate.

\subsection{Implications for Disclosure and Tool Design}
Gai and colleagues~\cite{gai2025} warn that mandatory AI disclosure may harm
marginalized groups. 
Our findings reinforce their broader caution about displaying author attributes, 
but suggest that disclosure effects depend on the population and on how
AI use is shown. In the AI-accustomed population we studied, an embedded
commit-message disclosure produced no detectable penalty. The seniority
label, however, did bias ratings in the same participants reviewing the
same code.
Status-based bias has been documented for decades~\cite{berger1972,nisbett1977},
and many review tools display author attributes such as seniority by
default. The design question is therefore not only ``should AI use be
shown?'' but ``which author attributes, if any, should the review tool
surface at all?''~\cite{murphyhill2024anonymous}

\subsection{Limitations}
Our sample is drawn from one large U.S.\ technology company with high
AI comfort; the null AI effect is a scope condition for an AI disclosure
embedded in the commit message, not a rebuttal of AI penalties
in less AI-comfortable populations. The seniority
manipulation contrasted two points on the level ladder (Level 1 vs.\ Principal); 
intermediate gradations may show smaller
effects. Vignette experiments lack the interactional dynamics of real
review, which may either amplify or attenuate status effects.

The AI disclosure was not visually highlighted and appeared alongside the
code review, but it occupied a substantial portion of the short commit
message; we therefore cannot characterize its overall salience
unambiguously. Because we did not include a participant-level manipulation
check for AI disclosure, we cannot determine how consistently participants
noticed it; failures to notice it would attenuate any AI-label effect.

Our design also cannot speak to whether AI disclosure interacts with
author gender, as Gai and colleagues~\cite{gai2025} report: we held
author gender ambiguous to isolate the AI and seniority cues. While we
group seniority and gender together as status cues, the
two are not interchangeable -- gender is a protected attribute with a
distinct legal and historical context, and a junior penalty does not
imply a gender penalty of a similar magnitude.


Finally, our AI disclosure was a free-text sentence appended to the
commit message, written before tooling converged on the structured
\texttt{Co-Authored-By:} trailer that Claude Code and other agents now
emit by default.
The default messages may induce different reviewer reactions than our disclosure. 

\section{Conclusion}
With an AI-use disclosure embedded in the commit message, disclosed AI use
produced no detected evaluation penalty in a population where AI use has
been normalized and AI was framed as tool-use assistance rather than autonomous
authorship. This result extends prior work by documenting a context in which
the AI-disclosure penalty does not appear. Such contexts may become increasingly
common as organizational AI adoption grows. In the same population, disclosed author seniority
drove a junior penalty on identical code that lowered both reviewers'
ratings of the code and their ratings of the author's competence and
effort. Bias from displayed author information remains consequential in
code review, and tool designers should consider which author attributes
review interfaces surface.

\section*{Acknowledgments}

Thanks to the participants who took the time to complete the study,
without whom this research would not have been possible. 
We also thank Barrett Amos, David Speirs, and 
the anonymous reviewers for their helpful feedback and suggestions.

\IEEEtriggeratref{8}
\bibliographystyle{IEEEtran}
\bibliography{references}

%
%

\clearpage

\appendices

\section{Robustness Checks}
\label{app:robustness}

\subsection{AI-disclosure result}

\paragraph{Evidence that participants noticed the disclosure.}
We did not include a formal AI manipulation check, so we searched the
optional per-review free-text feedback for spontaneous AI mentions. A
response counted as an AI mention if a case-insensitive regular
expression matched any of the following terms at word boundaries:
\texttt{AI}, \texttt{A.I.}, \texttt{Copilot}, \texttt{ChatGPT},
\texttt{Chat GPT}, \texttt{GPT}, \texttt{LLM}, \texttt{GenAI},
\texttt{Gen AI}, \texttt{generative}, \texttt{artificial intelligence},
\texttt{machine-generated}, or \texttt{auto-generated}; the last two
also matched a space in place of the hyphen. Each nonempty response
counted at most once. Among nonempty responses, the search matched
\AIMentionLabeled\ of \AIMentionLabeledNonempty\ AI-labeled reviews and
\AIMentionUnlabeled\ of \AIMentionUnlabeledNonempty\ non-AI-labeled
reviews (Fisher's exact OR=\AIMentionOR, \(\AIMentionP\)). This is
indirect evidence that at least some participants noticed the
disclosure, not a substitute for a participant-level manipulation check.

\paragraph{Power and prior effect sizes.}
A post-hoc power analysis indicates that the design had
\(\PowerWithin\%\) power to detect Gai and colleagues' \(d=\GaiD\) on
\textsc{CoderCompetence}. Translated into our $z$-composite units,
their effect would be approximately \(\GaiDInOurUnits\). Our lower 95\%
confidence bound (\(\AICompCIlo\)) excludes a penalty of that magnitude.

\paragraph{Excluding the AI-generated snippet.}
The AI result did not hinge on the AI-generated TicTacToe snippet,
which had no seeded bugs. Refitting the models to the \NExclReviews\
reviews of the three literature-derived snippets yielded no AI effect
on \textsc{CoderCompetence} (\(b=\ExclAICompCoef\),
\(\ExclAICompP\)), \textsc{CodeEffectiveness}
(\(b=\ExclAIEffCoef\), \(\ExclAIEffP\)), or
\textsc{CoderLaziness} (\(b=\ExclAILazyCoef\),
\(\ExclAILazyP\)).

\subsection{Author-seniority result}

\paragraph{Excluding the AI-generated snippet.}
After excluding TicTacToe, the seniority effect remained significant
on \textsc{CoderCompetence} (\(\ExclSrCompP\)) and
\textsc{CodeEffectiveness} (\(\ExclSrEffP\)).

\paragraph{Reviewer seniority.}
To test whether in-group favoritism drove the junior penalty, we refit
each DV with a significant seniority effect using
\(\mathrm{isSenior} \times \mathrm{usedAI} \times
\mathrm{reviewerIsSenior}\). The
\textsf{isSenior}\(\times\)\textsf{reviewerIsSenior} interaction was
non-significant across DVs, while the \textsf{isSenior} main effect
remained significant on \textsc{CoderCompetence}
(\(\RobustSrCompP\)), \textsc{CodeEffectiveness}
(\(\RobustSrEffP\)), and the individual code ratings. Code
Readability, Competent, and Gives Up Easily fell below the
\(p<0.05\) threshold only in this saturated three-way model, with
little change in their point estimates, consistent with reduced power
rather than effect instability.

\paragraph{Comment-requirement regimes.}
The comment-count null held when the optional- and required-comment
regimes were modeled separately: optional-comment reviews
(\(\CommentOptionalP\)) and required-comment reviews
(\(\CommentRequiredP\)).

\section{Field Comparison: Calibration and Caveats}
\label{app:decompose}

This section gives the calculation behind the $33\%$ figure cited in
Section~\ref{sec:decompose}.

\paragraph{Level mapping.} Murphy-Hill and colleagues' field result is
reported as an odds ratio of $0.38$ for a Staff (L6) author relative
to a SWE~II (L3) author---equivalently, an L3 author faces
$1/0.38\approx 2.6\times$ the odds of high pushback as an L6 author
on a comparable change. For the comparison to our experiment to be
meaningful, our Level 1-vs.-Principal contrast must span roughly the
same range as their L3-vs-L6 contrast. Cross-company level
mappings\footnote{See, e.g., the Google--\OurOrg{} alignment at
\url{https://www.levels.fyi}.} place \OurOrg's ``Software Engineer
I'' near Google L3 and \OurOrg's ``Principal Engineer'' near Google
L6, making the two contrasts roughly comparable in span.

\paragraph{Putting the two effects on a common scale.}
Pushback is a binary behavioral outcome and our DVs are continuous
ratings, so the units are not directly comparable on their face. The
standard meta-analytic move is to convert the odds ratio to a
Cohen's $d$ on the latent propensity that the binary indicator
thresholds: \(d = \log(\mathrm{OR})\cdot\sqrt{3}/\pi\)~\cite{sanchezmeca2003}.
For L3 vs.\ L6, that gives \(\log(0.38)\cdot\sqrt{3}/\pi \approx -0.534\):
the senior label is worth roughly $0.53$~SD less pushback in the
field. Our experiment moves \textsc{CoderCompetence} by
\(\SrCompCoef\)~SD when the title flips from Level 1 to Principal.
(Because \textsc{CoderCompetence} is $z$-scored before fitting and
\textsf{isSenior} is binary, this coefficient is itself a
standardized mean difference, on the same SD-of-the-DV scale as the
converted field estimate.) Comparing the two,
\(\SrCompCoef/0.534 \approx 33\%\).

\paragraph{Caveats.} The two SDs are SDs of different latent
constructs, the field model still includes residual confounding our
experiment sidesteps, and the conversion assumes the standard
logistic latent-variable model. We therefore treat $33\%$ as an
order-of-magnitude estimate, not a point estimate.

\section{Manipulation Check Reinterpreted}
\label{app:manip}

After each review, participants were also asked, ``Looking at the
commit message information, what is the level of the author of this
code?'' We included this as an attention check on the seniority
manipulation, but in practice participants appear to have answered it
as a competence judgment, not a recall task: only \MCPassRate\% of
responses matched the assigned label, but the responses themselves
correlated more strongly with code-quality ratings
(\(\rho=\MCRhoObs\), \(p<.001\)) than the assigned label did
(\(\rho=\MCRhoExp\), \(p<.001\)). The excess correlation
(\(\Delta\rho=\MCRhoDiff\)) is consistent with reviewers using the
level cue to form an opinion of the coder rather than passively
echoing it.
This interpretation is, in retrospect, supported by one of the pilots, 
where one developer mentioned this question seemed 
``too easy'' so they assumed they were supposed to answer with what 
level they \textit{thought} the developer was based on the code. 

\section{Survey Instrument}
\label{app:survey}

This section reproduces the full survey instrument as seen by participants.
The instrument was delivered as a single-page web application consisting of
five screens, described in the five subsections below. A persistent banner at
the top of every page after the consent screen read:

\begin{quote}
\textit{Your work won't be saved if you navigate away, go back, or refresh.}
\end{quote}

\subsection{Screen 1: Welcome and Consent}
\label{app:welcome}

\paragraph{Title.} Welcome to a study of code reviews

\paragraph{Project Title.} Examining the thought processes of Code Review

\paragraph{Co-Investigators.} \AnonAuthors

\paragraph{Invitation.} Thank you for taking the time to consider volunteering
in a research project at \OurOrg{}. The purpose of this research is to learn how
people think about and approach code reviews.

\paragraph{Procedures.} If you choose to participate, you will see a set of
questions asking about your experience with code reviews. You will then have
the opportunity to participate in a code review session. In this session, you
will be asked to review 4 pieces of code from 4 different engineers. After you
review the code, you will be asked to rate the code as well as your views on
the author of the code. After you are done reviewing the code, you will be
asked to answer a few questions about your experience. Finally, you will be
asked to email your results to the study coordinator. The entire session will
take about 30 minutes.

During the study, we plan to collect information about you such as your views
on code reviews and your thoughts on various code snippets. Additionally, we
will collect your email address if you choose to enter the sweepstakes. Your
email address will be stored separately from your responses, and the key
linking the two will be deleted as soon as your address and data is recorded
(likely within 3 weeks of data submission).

By participating in this study, you have the option to be entered into a
sweepstakes to win 1 of 4 \$50 gift cards. Your approximate chances of winning
depend on the total number of entries, but are no less than 1 in 250.

\paragraph{Benefits.} A benefit of participating in this study is that you
will practice your code review skills. Additionally, you will be contributing
to research that helps us understand developer experiences during code review.
We hope that findings from this research will help everyone involved in code
reviews to have an equitable experience. Lastly, you will have the option of
being entered into a drawing for 1 of 4 \$50 gift cards.

\paragraph{Risks.} The risks of participating are similar to what you might
experience while performing everyday tasks. A risk of participating in this
study is that you may feel uncomfortable answering questions about your
experience with code reviews. If you feel uncomfortable at any time, you may
stop participating in the study.

\paragraph{Considerations for Participants in People's Republic of China
(PRC).} All collected data will be stored in \OurStorage{} in the United States,
and not necessarily retained in the country of the participant.

\paragraph{Privacy and Confidentiality.} Researchers will keep your
participation and the information you share as confidential as possible. The
information you share will be labeled in our records with a code instead of
your name or other direct identifier. The key to this code will be stored
separately and destroyed after publication of the paper.

\paragraph{Data Sharing Plans.} We might share de-identified data from this
study with additional researchers who help with the study. They will retain
access to this data for 12 weeks.

\paragraph{Future Sharing Plans.} Researchers may share the results of this
study publicly, such as in journal articles or conference presentations, but
your name will not be included. Information and data collected from you during
this study may be used for future research studies or to improve products or
services at \OurOrg{}. If that happens, researchers will remove any direct
identifiers, like your name or email address, before sharing.

\paragraph{Participation is Your Choice.} Whether or not you participate is
entirely up to you. You can decide to participate now and stop participating
later.

\paragraph{Consent items.}
\begin{itemize}[leftmargin=*]
  \item \textbf{Consent checkbox (required):} I consent to the above and will
        take part in the study.
  \item \textbf{Sweepstakes participation (required radio):}
        \begin{itemize}
          \item Yes, I would like to be included in the sweepstakes
          \item No, I do not want to be included in the sweepstakes
        \end{itemize}
\end{itemize}

\subsection{Screen 2: Pre-Review Survey}
\label{app:presurvey}

After introducing what to expect (brief survey, four short code reviews with
per-review post-survey, closing survey), participants answered the following
four questions, each a required single-choice radio.

\begin{enumerate}[leftmargin=*]
  \item How often do you typically perform code reviews?
        \textit{Options:} Multiple times a day; Daily; A few times a week;
        Weekly; Every few weeks.
  \item How confident are you in your ability to conduct a thorough code
        review? \textit{Options:} Not at all confident; Slightly confident;
        Moderately confident; Very confident; Extremely confident.
  \item When reviewing code, how much emphasis do you put on the code's
        readability? \textit{Options:} Very low emphasis; Low emphasis;
        Neither high nor low emphasis; High emphasis; Very high emphasis.
  \item When reviewing code, how much emphasis do you put on the code's
        security? \textit{Options:} Very low emphasis; Low emphasis; Neither
        high nor low emphasis; High emphasis; Very high emphasis.
\end{enumerate}

\subsection{Screen 3: Code Review Intro}
\label{app:reviewintro}

\begin{quote}
\textbf{Upcoming Code Reviews.} You will now see \textbf{4 different code
reviews}. Before each review, you'll be shown who submitted it --- including
their name, photo, and role. Then you'll be asked to provide your feedback on
the code they submitted. Please review each submission thoughtfully.
\end{quote}

\subsection{Screen 4: Code Review (repeated four times)}
\label{app:codereview}

\subsubsection{Manipulation and randomization}
\label{app:randomization}

The study manipulates two between-snippet factors (author seniority and AI
disclosure), both randomized within-subject:

\begin{itemize}[leftmargin=*]
  \item \textbf{Reviewer pool (fixed).} Four fictional authors were presented
        with a non-gendered profile image and one of two titles:
        \begin{itemize}
          \item Reese --- Level~1 Engineer
          \item Skylar --- Level~1 Engineer
          \item Leslie --- Principal Engineer
          \item Pat --- Principal Engineer
        \end{itemize}
  \item \textbf{\textit{RV1: AI-use assignment.}} Within each seniority pair,
        exactly one author was randomly assigned \texttt{usedAI = true} and
        the other \texttt{usedAI = false}:
        \begin{itemize}
          \item With probability $0.5$: Reese used AI, Skylar did not;
                otherwise Skylar used AI, Reese did not.
          \item With probability $0.5$: Leslie used AI, Pat did not;
                otherwise Pat used AI, Leslie did not.
        \end{itemize}
        Every participant therefore saw exactly two AI-labeled authors and two
        non-AI-labeled authors, one of each at each seniority level.
  \item \textbf{\textit{RV2: Reviewer presentation order.}} The four reviewers
        were presented in a uniformly random permutation.
  \item \textbf{\textit{RV3: Code-snippet assignment.}} The four code snippets
        (Sections \ref{app:snippet-player}--\ref{app:snippet-addition}) were
        independently uniformly shuffled, then paired positionally with the
        shuffled reviewer list. Each reviewer is therefore associated with a
        single snippet, and the reviewer--snippet pairing is independent of
        the AI-use and seniority assignments.
\end{itemize}

The AI-use manipulation was surfaced to the participant only through the
per-review commit-message line (below); no icons, badges, or other markers
were shown. Photos and profile layout were identical across authors.

\subsubsection{Per-review screen content}

For each of the four reviews, the page contained:

\begin{enumerate}[leftmargin=*]
  \item A sticky header with the assigned author's name, title, and profile
        photo.
  \item A \emph{Commit message} block containing the snippet's
        \texttt{DESCRIPTION} (see Sections
        \ref{app:snippet-player}--\ref{app:snippet-addition}) followed by
        exactly one of the following sentences, depending on the random
        assignment \textit{RV1}:
        \begin{itemize}
          \item If \texttt{usedAI = true}: \textit{AI was used to assist with
                code generation.}
          \item If \texttt{usedAI = false}: \textit{No AI was used in
                generating the code.}
        \end{itemize}
  \item The instruction: \textit{How to comment: Hover over any line and
        click the bubble to leave feedback.}
  \item A side-by-side (\texttt{BEFORE}/\texttt{AFTER}) diff of the snippet.
      Participants could leave inline comments on any line of either side;
      comment submission was initially optional, and at least one comment was
      required for participants who submitted later in data collection.
  \item The post-review survey (below).
\end{enumerate}

\subsubsection{Post-review survey (per snippet)}

All items below are required except the final free-text item.

\begin{enumerate}[leftmargin=*]
  \item Looking at the commit message information, what is the level of the
        author of this code? \textit{Options:} Level~1; Level~2; Senior;
        Principal.
  \item Overall, how would you rate the quality of this code?
        \textit{Options:} Very poor; Poor; Mediocre; Good; Excellent.
  \item How would you rate the readability of this code?
        \textit{Options:} Very poor; Poor; Mediocre; Good; Excellent.
  \item How would you rate the security of this code?
        \textit{Options:} Very poor; Poor; Mediocre; Good; Excellent.
  \item How would you rate coding capabilities of this author?
        \textit{Options:} Very poor; Poor; Mediocre; Good; Excellent.
  \item Please rate the extent to which you believe the following adjectives
        or phrases accurately describe the author of this code. Each
        characteristic was rated on a 7-point Likert scale (Strongly
        disagree; Disagree; Somewhat disagree; Neither agree nor disagree;
        Somewhat agree; Agree; Strongly agree):
        \begin{itemize}
          \item Competent
          \item Irresponsible
          \item Skillful
          \item Lazy
          \item Gives up easily
          \item Masterful
        \end{itemize}
  \item Imagine you are a hiring manager responsible for recruiting a software
        engineer at your company. Based on the code you just reviewed, how
        likely is it that you would recommend the author to be considered for
        the position? \textit{Options:} Very unlikely; Unlikely; Neither
        likely nor unlikely; Likely; Very likely.
  \item Please provide any other feedback (\emph{Optional}, free text).
\end{enumerate}

\subsection{Screen 5: Closing Survey}
\label{app:closingsurvey}

All items below are required except the two optional free-text fields at the
end.

\begin{enumerate}[leftmargin=*]
  \item What is your current job level? \textit{Options:} Level~1; Level~2;
        Senior; Principal or above.
  \item What is your gender identity? \textit{Options:} Woman; Man;
        Non-binary; Prefer not to answer; Prefer to self-describe
        (free-text field required if this option is chosen).
  \item How often do you use AI tools (like GitHub Copilot, ChatGPT, etc.)
        when \textbf{conducting code reviews}? \textit{Options:} Never;
        Occasionally; Sometimes; Often; Always.
  \item \textbf{Likert block (code reviews).} Please rate the extent to which
        you agree or disagree with the following statements about using AI
        for code reviews. Each statement was rated on a 7-point scale
        (Strongly disagree; Disagree; Somewhat disagree; Neither agree nor
        disagree; Somewhat agree; Agree; Strongly agree):
        \begin{itemize}
          \item I would be comfortable with my manager knowing that I am
                using AI for code reviews.
          \item I would be comfortable with my colleagues knowing that I am
                using AI for code reviews.
          \item I would want my manager to know I was using AI for code
                reviews.
          \item I would want my colleagues to know I was using AI for code
                reviews.
        \end{itemize}
  \item How often do you use AI tools (like GitHub Copilot, ChatGPT, etc.)
        when \textbf{generating new code}? \textit{Options:} Never;
        Occasionally; Sometimes; Often; Always.
  \item \textbf{Likert block (code generation).} Please rate the extent to
        which you agree or disagree with the following statements about using
        AI for code generation (same 7-point scale as above):
        \begin{itemize}
          \item I would be comfortable with my manager knowing that I am
                using AI for code generation.
          \item I would be comfortable with my colleagues knowing that I am
                using AI for code generation.
          \item I would want my manager to know I was using AI for code
                generation.
          \item I would want my colleagues to know I was using AI for code
                generation.
        \end{itemize}
  \item \textbf{AI automation comfort.}
        How comfortable would you feel if an
        AI system wrote, reviewed code, and checked it in --- without any
        human oversight --- in the following situations? Each situation was
        rated on a 6-point scale (Very uncomfortable; Somewhat uncomfortable;
        Neutral; Somewhat comfortable; Very comfortable; NA):
        \begin{itemize}
          \item Remove unused (``dead'') code
          \item Updating dependency or package versions to the latest minor
                release
          \item Refactoring code for style or linting compliance
          \item Renaming variables for clarity (no functional change)
          \item Removing old no longer relevant comments / commented-out code
          \item Fixing small, well-defined bugs (e.g., off-by-one errors,
                null pointer exceptions)
          \item Removing feature flags or flighting gates after a rollout is
                complete
          \item Adding unit tests for existing code
          \item Implementing new features
          \item Changing business logic
          \item Making architectural changes (e.g., moving components to
                microservices)
        \end{itemize}
  \item Do you have any thoughts on fully automated code development? For
        example, reasons for or against, situations where it might be safe,
        how you would feel about it, etc. (\emph{Optional}, free text.)
  \item Any additional comments about AI usage in code reviews?
        (\emph{Optional}, free text.)
\end{enumerate}

\section{Code Snippets Shown to Participants}
\label{app:snippets}

Each of the four snippets below was presented as a side-by-side
\texttt{BEFORE}/\texttt{AFTER} diff alongside the given commit message.
Snippets spanning multiple files are separated by \texttt{FILENAME} markers.
Snippet-to-author pairing was randomized per participant
(Section~\ref{app:randomization}, \textit{RV3}). Long source lines are
soft-wrapped (continuation marked with $\hookrightarrow$).

\subsection{Snippet ``PlayerRegistration''}
\label{app:snippet-player}

\paragraph{Commit message.} Refactor player registration method to accept
board parameter and prevent duplicate registrations. Also adding test
coverage.

\paragraph{Before (main file).}\mbox{}\\ \textit{(The web diff rendered 10 blank
lines of padding between the signature and body to align with the
\texttt{AFTER} side; omitted here.)}
\begin{lstlisting}[language=Java]
    public void registerPlayer(Player p) {
        this.players.add(p);
        Square square = this.startSquares.get(this.startSquareIndex);

        // TODO put player on square

        this.startSquareIndex++;
        this.startSquareIndex %= this.startSquares.size();
    }
\end{lstlisting}

\paragraph{After (main file).}\mbox{}\\
\begin{lstlisting}[language=Java]
    public void registerPlayer(Player p, Board b) {
        if (this.players == null){
            this.players = new ArrayList<Player>();
        }

        this.board = b;
        this.inProgress = false;

        if (this.players.contains(p)) {
            return;
        }

        this.players.add(p);
        Square square = this.startSquares.get(this.startSquareIndex);

        p.occupy(square);

        this.startSquareIndex++;
        this.startSquareIndex %= this.startSquares.size();
    }
\end{lstlisting}

\paragraph{After (\texttt{LevelTest.java}; \texttt{BEFORE} was empty).}\mbox{}\\
\begin{lstlisting}[language=Java]
Board board = mock(Board.class);

@Test
public void registerPlayer() {
    Square square1 = mock(Square.class);
    Player p = mock(Player.class);
    level.registerPlayer(p, board);
    verify(p).occupy(square1);
}

@Test
public void registerPlayerTwice() {
    Square square1 = mock(Square.class);
    Player p = mock(Player.class);
    level.registerPlayer(p, board);
    level.registerPlayer(p, board);
    verify(p, times(1)).occupy(square1);
}

@Test
public void registerSecondPlayer() {
    Square square2 = mock(Square.class);
    Player p1 = mock(Player.class);
    Player p2 = mock(Player.class);
    level.registerPlayer(p1, board);
    level.registerPlayer(p2, board);
    verify(p2).occupy(square2);
}

@Test
public void registerNullPlayer() {
    Square square2 = mock(Square.class);
    Player p1 = null;
    level.registerPlayer(p1, board);
    verify(p1, times(1)).occupy(square1);
}

@Test
public void registerThirdPlayer() {
    Square square1 = mock(Square.class);
    Player p1 = mock(Player.class);
    Player p2 = mock(Player.class);
    Player p3 = mock(Player.class);
    level.registerPlayer(p1, board);
    level.registerPlayer(p2, board);
    level.registerPlayer(p3, board);
    verify(p3).occupy(square1);
}
\end{lstlisting}

\subsection{Snippet ``TicTacToe''}
\label{app:snippet-ttt}

\paragraph{Commit message.} This pull request is for a tic-tac-toe winner
detection function. Updating to fix bug.

\paragraph{Before (main file).}\mbox{}\\
\begin{lstlisting}[language=Python]
def check_winner(board):
    # Check rows and columns
    for i in range(3):
        if all(cell == board[i][0] and cell != " " for cell in board[i]):
            return board[i][0]
        if all(board[j][i] == board[0][i] and board[j][i] != " " for j in range(3)):
            return board[0][i]

    # Check diagonals
    if board[0][0] == board[1][1] == board[2][2] != " ":
        return board[0][0]
    if board[0][2] == board[1][1] == board[2][2] != " ":
        return board[0][2]

    return None
\end{lstlisting}

\paragraph{After (main file).}\mbox{}\\
\begin{lstlisting}[language=Python]
def check_winner(board):
    # Check rows and columns
    for i in range(3):
        if all(cell == board[i][0] and cell != " " for cell in board[i]):
            return board[i][0]
        if all(board[j][i] == board[0][i] and board[j][i] != " " for j in range(3)):
            return board[0][i]

    # Check diagonals
    if board[0][0] == board[1][1] == board[2][2] != " ":
        return board[0][0]
    if board[0][2] == board[1][1] == board[2][0] != " ":
        return board[0][2]

    return None
\end{lstlisting}

\paragraph{After (\texttt{test\_tictactoe.py}; \texttt{BEFORE} was empty).}\mbox{}\\
\begin{lstlisting}[language=Python]
import unittest
from unittest.mock import patch, MagicMock

class TestTicTacToe(unittest.TestCase):

    def setUp(self):
        self.board = [[" " for _ in range(3)] for _ in range(3)]

    def test_check_winner_row(self):
        self.board[0] = ["X", "X", "X"]
        self.assertEqual(check_winner(self.board), "X")

    def test_check_winner_column(self):
        for i in range(3):
            self.board[i][0] = "O"
        self.assertEqual(check_winner(self.board), "O")

    def test_check_winner_diagonal(self):
        for i in range(3):
            self.board[i][i] = "X"
        self.assertEqual(check_winner(self.board), "X")

    def test_check_winner_anti_diagonal(self):
        self.board[0][2] = "O"
        self.board[1][1] = "O"
        self.board[2][0] = "O"
        self.assertEqual(check_winner(self.board), "O")

    def test_no_winner(self):
        self.assertIsNone(check_winner(self.board))
\end{lstlisting}

\subsection{Snippet ``BigBox''}
\label{app:snippet-bigbox}

\paragraph{Commit message.} Implement box calculation algorithm to optimize
big box usage before small boxes with comprehensive test coverage.

\paragraph{Before (main file).}\mbox{}\\
\begin{lstlisting}[language=Java]
public class ExerciseBoxes {
    /*
    Given a total number of boxes, we have to return the
    number of small boxes we need to use to reach the total
    number of boxes, with the constraint that we should first use
    as many big boxes as possible.

    1 big box = 5 small boxes

    For example, if we have: total = 16, big = 3, small = 5, the
    result would be 1. ( 3 big boxes = 15 boxes, so only 1 small box is
    needed).
    Return -1 in the case in which there are not enough small boxes
     */
}
\end{lstlisting}

\paragraph{After (main file).}\mbox{}\\
\begin{lstlisting}[language=Java]
public class ExerciseBoxes {
    /*
    Given a total number of boxes, we have to return the
    number of small boxes we need to use to reach the total
    number of boxes, with the constraint that we should first use
    as many big boxes as possible.

    1 big box = 5 small boxes

    For example, if we have: total = 16, big = 3, small = 5, the
    result would be 1. ( 3 big boxes = 15 boxes, so only 1 small box is
    needed).
    Return -1 in the case in which there are not enough small boxes
     */
    public Integer calculate(Integer small, Integer big, Integer total) {
        Integer maxBigBoxes = total / 5;
        Integer bigBoxesWeCanUse;

        if (maxBigBoxes < big){
            bigBoxesWeCanUse = maxBigBoxes;
        } else {
            bigBoxesWeCanUse = big;
        }

        total = total - (bigBoxesWeCanUse * 5);

        if(small <= total)
            return -1;

        return total;
    }
}
\end{lstlisting}

\paragraph{Before (\texttt{ExerciseBoxesTest.java}).}\mbox{}\\
\begin{lstlisting}[language=Java]
import org.junit.Test;

import static org.junit.Assert.assertEquals;
public class ExerciseBoxesTest {
}
\end{lstlisting}

\paragraph{After (\texttt{ExerciseBoxesTest.java}).}\mbox{}\\
\begin{lstlisting}[language=Java]
import org.junit.Test;

import static org.junit.Assert.assertEquals;
public class ExerciseBoxesTest {
    @Test
    public void simpleTest(){
        ExerciseBoxes ex = new ExerciseBoxes();
        Integer expected = 5;
        Integer result = ex.calculate(7,3,20);
        assertEquals(expected, result);
    }

    @Test
    public void impossibleTest(){
        ExerciseBoxes ex = new ExerciseBoxes();
        Integer expected = -1;
        Integer result = ex.calculate(3,3,20);
        assertEquals(expected, result);
    }
}
\end{lstlisting}

\subsection{Snippet ``Addition''}
\label{app:snippet-addition}

\paragraph{Commit message.} Implement array-based number addition algorithm
with carry handling and comprehensive test coverage.

\paragraph{Before (main file).}\mbox{}\\
\begin{lstlisting}[language=Java]
public class ExerciseSumArray {
    /*
    Given 2 Lists representing numbers (e.g., [3,4] = 34, [9,8] = 98),
    calculate the sum of 2 Lists, and return the result in an List.
    For example:
    [1, 0, 0] + [4,0] = [1,4,0]
    [6,7] + [0] = [6,7]
     */
}
\end{lstlisting}

\paragraph{After (main file).}\mbox{}\\
\begin{lstlisting}[language=Java]
public class ExerciseSumArray {
    /*
    Given 2 Lists representing numbers (e.g., [3,4] = 34, [9,8] = 98),
    calculate the sum of 2 Lists, and return the result in an List.
    For example:
    [1, 0, 0] + [4,0] = [1,4,0]
    [6,7] + [0] = [6,7]
     */
    public ArrayList<Integer> getSum(List<Integer> firstNumber, List<Integer> secondNumber ){
        ArrayList<Integer> result = new ArrayList<Integer>();

        int carry = 0;
        Collections.reverse(firstNumber);
        Collections.reverse(secondNumber);

        for (int i = 0; (i < Math.max(firstNumber.size(), secondNumber.size())); i ++){
            Integer firstValue = i < firstNumber.size() ? firstNumber.get(i) : null;
            Integer secondValue = i < secondNumber.size() ? secondNumber.get(i) : null;

            int res = firstValue + secondValue + carry;

            carry = 0;
            if (res > 10){
                carry = 1;
                res = res % 10;
            }
            result.add(res);
        }

        if (carry >= 0)
            result.add(carry);

        Collections.reverse(result);
        return result;
    }
}
\end{lstlisting}

\paragraph{Before (\texttt{ExerciseSumArrayTest.java}).}\mbox{}\\
\begin{lstlisting}[language=Java]
public class ExerciseSumArrayTests {
}
\end{lstlisting}

\paragraph{After (\texttt{ExerciseSumArrayTest.java}).}\mbox{}\\
\begin{lstlisting}[language=Java]
public class ExerciseSumArrayTests {
    @Test
    public void simpleSumTest(){
        ExerciseSumArray ex = new ExerciseSumArray();
        List<Integer> first = Arrays.asList(6,8);
        List<Integer> second = Arrays.asList(5,0);
        List<Integer> result = Arrays.asList(1,1,8);

        assertEquals(result, ex.getSum(first, second));
    }

    @Test
    public void sumNumbersDifferentLengthTest(){
        ExerciseSumArray ex = new ExerciseSumArray();
        List<Integer> first = Arrays.asList(4,9);
        List<Integer> second = Arrays.asList(7,8);
        List<Integer> result = Arrays.asList(1,2,7);

        assertEquals(result, ex.getSum(first, second));
    }

    @Test
    public void sumZeroTest(){
        ExerciseSumArray ex = new ExerciseSumArray();
        List<Integer> first= Arrays.asList(6,7);
        List<Integer> second = Arrays.asList(5,5);
        List<Integer> result = Arrays.asList(1,2,2);

        assertEquals(result, ex.getSum(first, second));
    }
}
\end{lstlisting}

\end{document}